\documentclass[aps,pre,showpacs,reprint,superscriptaddress,floatfix]{revtex4-1}
\usepackage{graphicx,color,epsfig}
\usepackage{amssymb,amsmath,wasysym}
\begin{document}
 
\title{Rise of an alternative majority against opinion leaders}
\author{K. Tucci}
\affiliation{Grupo de Caos y Sistemas Complejos, Centro de F\'isica Fundamental, Universidad de Los Andes, M\'erida, Venezuela.}
\author{J. C. Gonz\'alez-Avella}
\affiliation{Departamento de F\'isica, Pontificia Universidade Cat\'olica do Rio de Janeiro, Caixa Postal 38071, 22452-970 RJ, Brazil.}
\author{M. G. Cosenza}
\affiliation{Grupo de Caos y Sistemas Complejos, Centro de F\'isica Fundamental, Universidad de Los Andes, M\'erida, Venezuela.}
\date{\today}

\begin{abstract}
We investigate the role of opinion leaders or influentials in the collective behavior of a social system.
Opinion leaders are characterized by their unidirectional influence on other agents. 
We employ a model based on Axelrod's dynamics for cultural interaction 
among social agents that allows for non-interacting states. 
We find three collective phases in the space of parameters of the system, given by the
fraction of opinion leaders and a quantity representing the number of available states: 
one ordered phase having the state imposed by the leaders;
another nontrivial ordered phase consisting of a majority group in a state orthogonal or alternative to that of the opinion leaders,
and a disordered phase, where many small groups coexist. 
We show that the spontaneous rise of an alternative group in the presence of opinion leaders 
depends on the existence of a minimum number of long-range connections in the underlying network. 
This phenomenon challenges the common idea that influentials are fundamental to propagation 
processes in society, such as the formation of public opinion.
 \end{abstract}
%\begin{keyword}
%Opinion leaders; Social dynamics; Alternative ordering. 
%\end{keyword}
 \maketitle

\section{Introduction}
Propagation processes describe many important activities in societies, 
such as  opinion formation, epidemic propagation, culture dissemination, viral marketing, and innovation diffusion, 
and their study is of much interest in social, biological and political sciences
\cite{Rogers,Granovetter,May,Hogg,Axelrod}.
A central argument in the research of these processes  
has been that the most influential agents  -- a minority of individuals who influence an exceptional number of their peers -- 
are fundamental to the propagation of behaviors in a society \cite{Katz,Valente,Roch}.
These agents are called opinion leaders, influentials, or spreaders \cite{Merton,Weiman,Lloyd,Yamir};
in opinion dynamics models they are also named zealots, inflexibles, or committed agents \cite{Mobilia,Verma,Galam,Xie}. 
Experimental investigation on a social network (Facebook) revealed that influential individuals are actually less susceptible 
to influence than noninfluential individuals \cite{Aral}. 
The activity of opinion leaders has been considered an important 
resource in the diffusion of information and marketing strategies in society \cite{Gladwell}. 
On the other hand, a common idea  in social networks has been that the most connected people 
are responsible for the largest scale of spreading processes \cite{Barabasi,Pastor,Cohen}. 

Recently, some works have questioned this so-called `influentials hypothesis' \cite{Watts}; for example,
it has been argued that social contagion is driven by a critical mass of individuals susceptible to influence rather 
than by influentials \cite{Watts}; and that there are circumstances under which the most highly connected or the most central people 
have little effect on a spreading process \cite{Stanley}. 

Although the notion of opinion leadership seems clear, 
precisely how and when the influence of opinion leaders over their 
environment shapes opinions and trends across
entire societies remains an open problem. 
In this paper we present a dynamical agent-based model to investigate  
the collective behavior of a social system under the influence of  
opinion leaders. 
We define opinion leaders as agents that can affect the state of other agents, but their state remains unchanged; i. e.,
we assume that the interaction leader-agent is unidirectional. This simplifying assumption expresses the basic nature of
the interaction with opinion leaders as described in the literature 
\cite{Katz,Valente,Roch,Merton,Weiman,Lloyd,Yamir,Mobilia,Verma,Galam,Xie,Aral,Gladwell}. 
This also corresponds to the notion of cultural status proposed by Axelrod \cite{Axelrod}.
The unidirectional interaction dynamics of opinion leaders is similar to that of an external field, or mass media, acting on
a social system \cite{JC1,JC3}.
Then, opinion leaders can be regarded as distributed mediators or originators of mass media messages 
in a society \cite{Katz,Valente,Roch}.

As interaction dynamics, we employ Axelrod's rules for the dissemination of culture among social agents \cite{Axelrod}, 
a non-equilibrium model that has attracted much attention from physicists \cite{Castellano,Klemm,JC1,JC3,Kuperman,Galla,Floria,Lim,Fede}.
In this model, the interaction rule between agents is such that no interaction
takes place for some relative values characterizing the states of the agents. This type of interaction is
common in social and biological systems where there is often some bound for the occurrence of interaction
between agents, such as a similarity condition for the state variable \cite{Deffuant,Mikhailov,Weis,Krause}.

We show that for low values of the fraction of opinion leaders present,
the system is driven towards the opinion state of the leaders. 
However,   
above some critical value of the fraction of opinion leaders, 
we find the  nontrivial 
result that a majority group emerges in the system possessing a state 
non-interacting -- or alternative -- with that of the leaders, challenging the influentials hypothesis.
When the number of available states for the agents is large, the system reaches a disordered state
where many small groups coexist.
These three collective phases are characterized on the space of parameters of the system, given by the
fraction of opinion leaders and a quantity representing the number of available states.

\section{Social dynamics in the presence of opinion leaders}
We consider a system of $N$ agents located at the nodes of a network.  
The agents are distributed into two populations: 
a population $\alpha$ representing opinion leaders having a fixed opinion or cultural state, with size $N_\alpha$; 
and a population $\beta$ of agents capable of changing their states, with size $N_\beta$, such that $N_\alpha +N_\beta=N$. 
The fraction of opinion leaders is $\rho=N_\alpha/N$.
Both opinion leaders and agents in $\beta$ are randomly assigned to the nodes in the network. 
The set of neighbors of an agent $i \in \beta$ is denoted by $\nu_i$. 
The state of agent $i \in \beta$ is given by an $F$-component vector $x_\beta^f(i)$, $(f=1,2,\ldots,F)$, where each component can take any 
of the $q$ different values $x_\beta^f(i) \in \{0,1,\ldots,q-1\}$.  
On the other hand, opinion leaders share the same state, i.e., if $i \in \alpha$, $x_\alpha^f(i)=y^f$, where each component
$y^f$ is fixed and remains invariant during the evolution of the system.  
At any given time, a selected agent in population $\beta$ 
can interact with any agent in its neighborhood, 
which can be either another agent in $\beta$ or an opinion leader in population $\alpha$, in each case following the dynamics of 
Axelrod's model for cultural influence \cite{Axelrod}. 
As initial condition, each state $x_\beta^f(i)$ is randomly assigned one of the $q^F$ possible vector states with a uniform probability. 
Then, the dynamics of the system is defined by the following iterative algorithm:
\begin{enumerate}
\item  Select at random an agent $i \in \beta$ and an agent $j \in \nu_i$ whose state we generically 
denote by $x^f(j)$.
 \item Calculate the overlap between the states of agents $i$ and $j$, defined as
\begin{equation}
  d(i,j)=    \left\lbrace 
 \begin{array}{ll}  
 \sum_{f=1}^F \delta_{x_\beta^f(i) \, y^f}, & \mbox{if $j \in \alpha$}, \\
\sum_{f=1}^F  \delta_{x_\beta^f(i)\,  x_\beta^f(j)}, & \mbox{if $j \in \beta$}.
  \end{array}
     \right.  
\end{equation}
\item If $0 < d(i,j) < F$, with probability $d(i,j)/F$ choose $h$ randomly such that $x_\beta^h(i) \neq x^h(j)$ and set 
$x_\beta^h(i)=y^h$ if $j \in \alpha$,  
or $x_\beta^h(i)=x_\beta^h(j)$ if $j \in \beta$. 
If $d(i,j)=0$ or $d(i,j)=1$, the state $x_\beta(i)$ does not change. 
\end{enumerate}

In this model, opinion leaders can affect the states of other agents, but their state remains unchanged.
Thus the dynamical changes of the system occur on the population $\beta$. 
We shall consider small values of $\rho$ to take into account 
the observation that opinion leaders constitute a minority in a social system \cite{Katz,Valente,Roch,Aral}.

When no opinion leaders are present $(\rho=0)$, a system subject to Axelrod's dynamics  
reaches a stationary configuration in any finite network, where the agents form domains of different sizes.
A domain is a set of connected agents that share the same state. A homogeneous phase in a system is characterized by $d(i,j)=F$, 
$\forall i,j$. 
The coexistence of several domains  corresponds to an
inhomogeneous or disordered phase in a system.
It is known that, on several networks, the system reaches a homogeneous phase for values $q < q_c$, and a disordered phase for $q > q_c$,
where $q_c$ is a critical point \cite{Castellano,Klemm}. 

We consider two order parameters to characterize the collective behavior of the system under the influence of opinion leaders:  
the normalized average size of the largest domain in population $\beta$, called $S_\beta$;
and the  normalized average size of the largest domain possessing the state of the opinion leaders in population $\beta$, 
denoted by $S_\beta(y^f)$.

First, we study the model in a fully connected network, where every agent in $\beta$ can interact with any other in the system; 
i. e., $\nu_i=\alpha \cup \beta, \forall i$. 
In this situation, the fraction of opinion leaders $\rho$ also represents the probability for the agent--opinion leader interactions.
For a fully connected network with Axelrod's dynamics, in absence of opinion leaders, 
the critical value $q_c$ depends on the system size as $q_c \sim N$ \cite{Fede}.

\begin{figure}[h]
\begin{center}
\includegraphics[width=0.335\linewidth,angle=90]{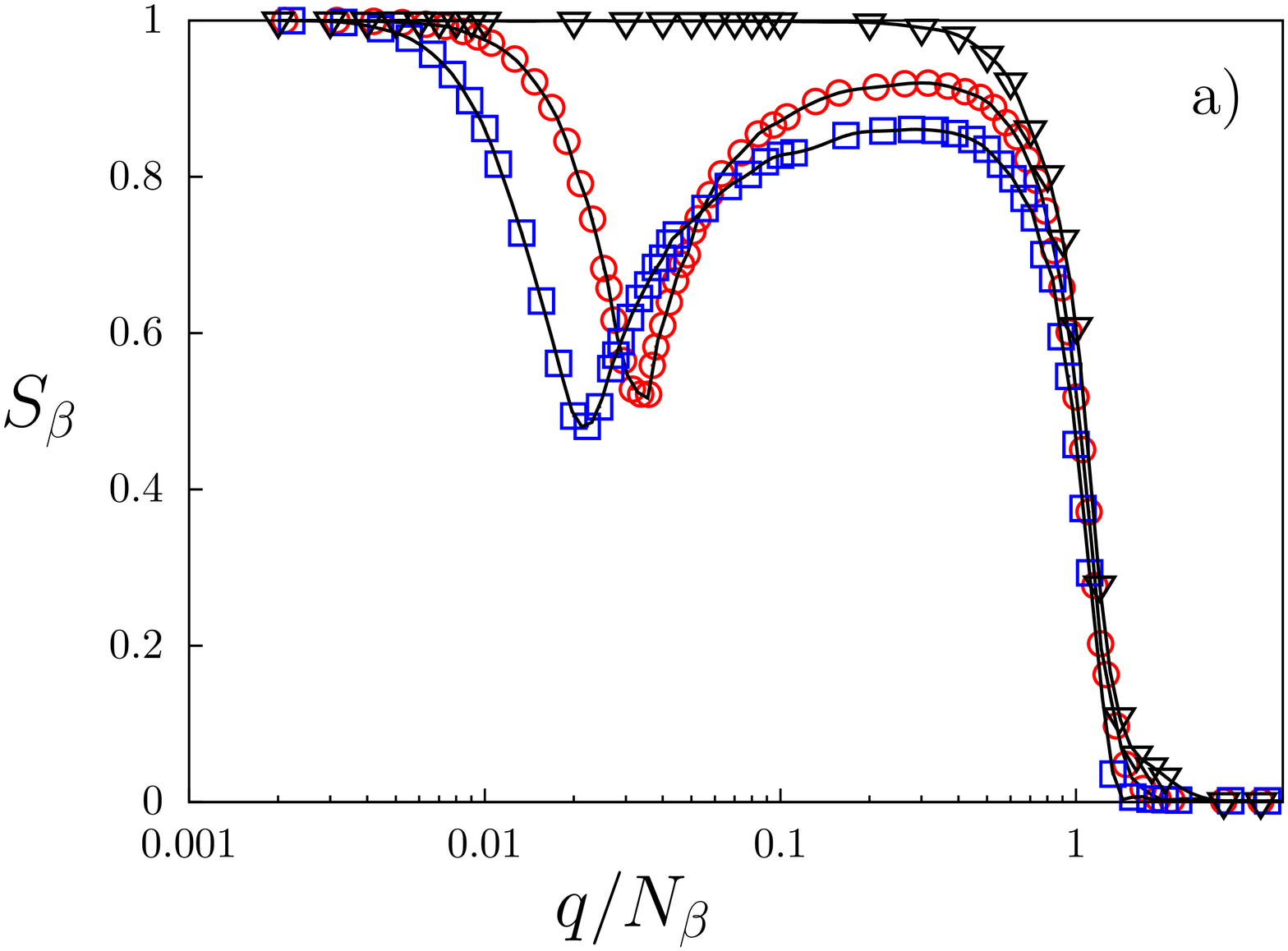}\\
\vspace{1mm}
\includegraphics[width=0.335\linewidth,angle=90]{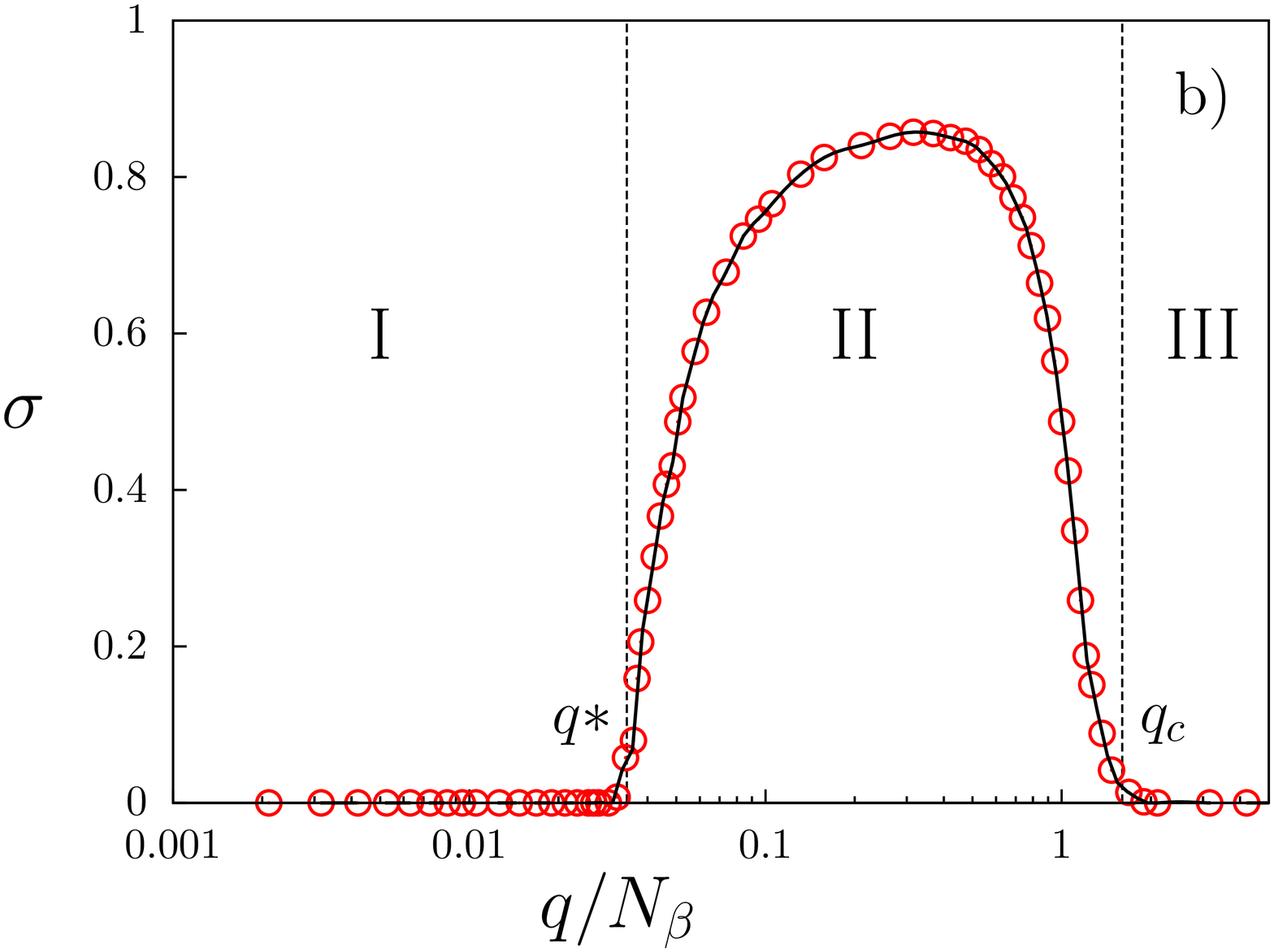}
\end{center}
\caption{a) Order parameter $S_\beta$ as a function of $q/N_\beta$ for parameter values
$\rho=0$ (triangles), $\rho=0.05$ (circles), and $\rho=0.1$ (squares).
b) Quantity $\sigma= S_\beta- S_\beta(y^f)$ as a function of $q/N_\beta$ for 
$\rho=0.05$ (circles). 
Labels I--III refer to the phases shown in the phase diagram of Fig.~(\ref{F3}). 
Fixed parameters are $N=1000$, $F=5$. 
Each data point is an average over $50$ independent realizations of initial conditions.}
\label{F1}   
\end{figure}

Figure~(\ref{F1}a) shows the order parameter  $S_\beta$ as a function 
of $q/N_\beta$ for different values of the fraction  $\rho$. 
When opinion leaders are absent ($\rho=0$), population $\beta$ reaches an
ordered phase for values $q < q_c$, characterized by $S_\beta \to 1$, and a disordered phase
for  $q > q_c$, for which $S_\beta \to 0$.
As $\rho$ increases, $S_\beta$ 
exhibits a local minimum at a value $q_*/N_\beta < q_c/N_\beta$ that depends on $\rho$.
Similarly, the value of $q_c$ scales as $q_c \sim N_\beta= N(1-\rho)$, as seen in Fig.~(\ref{F1}a).
To elucidate the origin of the local minimum, in Fig.~(\ref{F1}b) we show the quantity  $\sigma \equiv S_\beta-S_\beta(y^f)$ as a 
function of $q/N_\beta$, for a fixed value of $\rho$. For
$q < q_*$, the largest domain in $\beta$ acquires 
the state of the opinion leaders $y^f$, and therefore $\sigma= 0$. However, for $q_* < q <  q_c$, 
the largest domain corresponds to another state non-overlapping with $y^f$, i.e., $S_\beta > S_\beta(y^f)$, and  therefore $\sigma > 0$. 
For values $q > q_c$ disorder appears in the system, and both $S_\beta \to 0$, $S_\beta(y^f) \to 0$, and thus $\sigma=0$.

%%%%%%%%%%%%%%%%%%%%%%%%%%%%%%%%%%%%%%%%%%%%%%%%%%%%%%%%%%%%%%%%%%%%%%
%%%%%%%%%%%%%%%%%%%%%%%%%%%%%%%%%%%%%%%%%%%%%%%%%%%%%%%%%%%%%%%%%%%

\begin{figure}[h]
\begin{center}
\includegraphics[width=0.41\linewidth,angle=90]{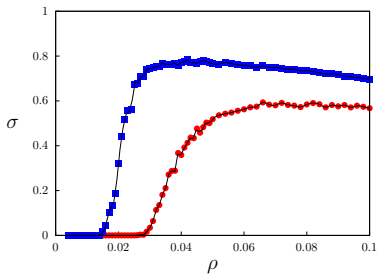}
\end{center}
\caption{ 
Quantity $\sigma= S_\beta- S_\beta(y^f)$ as a function of $\rho$ for  parameter values $q = 50$  (circles) and 
$q = 100$ (squares). Fixed parameters are $N=1000$, $F=5$. 
Each data point is an average over $50$ independent realizations of initial conditions.
}
\label{F2}      
\end{figure}

To investigate the role of the opinion leaders on the behavior of the largest domain in population $\beta$, 
in Fig.~(\ref{F2}) we show the dependence of  $\sigma$ on
the fraction $\rho$, for values of parameters $q_*<q < q_c$. 
As  $\rho$ increases, a competition ensues between the spontaneous order emerging in population
$\beta$ due to the inter-agent interactions and the order being imposed by the opinion leader-agent interaction in $\beta$.
For small values of  $\rho$, the largest domain in population $\beta$ is driven 
towards the state of the opinion leaders $y^f$,  and thus $\sigma=0$. 
Also as  $\rho$ increases, the size of population $\beta$ decreases and, as a consequence,
there are fewer agents in $\beta$ whose states share some features with the state  $y^f$.
There is a critical value of the fraction $\rho$ above which 
the largest domain forming in $\beta$ no longer converges to the state of the opinion leaders $y^f$, but reaches 
a state non-interacting with $y^f$, characterized by $\sigma > 0$. 
Thus, above some threshold value of the fraction of opinion leaders, their presence actually promotes the emergence of 
a majority group in population $\beta$ possessing a state orthogonal to that of the leaders.

%%%%%%%%%%%%%%%%%%%%%%%%%%%%%%%%%%%%%%%%%%%%%%%%%%%%%%%%%%%%%%%%%%%%%%
%%%%%%%%%%%%%%%%%%%%%%%%%%%%%%%%%%%%%%%%%%%%%%%%%%%%%%%%%%%%%%%%%%%%%

\begin{figure}[h]
\begin{center}
\includegraphics[width=0.41\linewidth,angle=90]{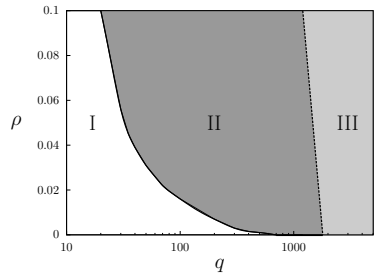}
\end{center}
\caption{Phase space on the plane $(q,\rho)$, with fixed parameters  $N=1000$, $F=5$.
Regions where phases I--III occur are indicated. The critical boundary between phases II and III
follows the relation $q_c \sim N(1-\rho)$, which we have verified for different system sizes.}
\label{F3}      
\end{figure}

In Fig.~(\ref{F3}) we show the collective behavior of the system subject to the action of opinion leaders on the space of parameters $(q,\rho)$. 
Three phases can be characterized on this space:
(I) an ordered phase imposed by the opinion leaders for $q < q_*$, for which $\sigma = 0$ and $S_\beta =S_\beta(y^f) \sim 1$; 
(II) an ordered phase in a state non-interacting with the state of the opinion leaders (i. e. the overlap between the
ordered state and $y^f$ is zero) for $q_* < q < q_c$, for which $\sigma >0$ and $S_\beta > S_\beta(y^f)$; and
(III) a disordered phase for $q > q_c$, for which  both $S_\beta \to 0$, $S_\beta(y^f) \to 0$, and $\sigma=0$.
In phase I, opinion leaders are successful at inducing their cultural state to the largest domain
formed in the system. Phase II corresponds to a situation where a group of agents 
in $\beta$ spontaneously orders in a cultural state non-interacting
with that being transmitted by the opinion leaders and emerges as the majority group. 
In a social context, phase II represents the rise of an alternative group against the imposition of a fixed cultural message 
or opinion by an influential group, or by a spatially distributed mass media message.

%%%%%%%%%%%%%%%%%%%%%%%%%%%%%%%%%%%%%%%%%%%%%%%%%%%%%%%%%%%%%%%%%%%%%%
%%%%%%%%%%%%%%%%%%%%%%%%%%%%%%%%%%%%%%%%%%%%%%%%%%%%%%%%%%%%%%%%%%%%%

\section{Network of interacting states.}
The collective dynamics of the system associated to each of these phases can be described in terms of the 
changes in the connectivity of the network of interacting states defined as follows: agents in the system are considered as the nodes 
of the network, and a link between any two nodes exists if their state variables share at least one component. 
This network of interacting states is, in general, different from the network of neighbors -- a fully connected network in the present case.
Let $\beta_0$ be the subset of agents in $\beta$ whose state vector initially shares at least one component with 
the state of the opinion leaders $y^f$.  This subset has size $N_{\beta_0}=N(1-\rho)[1 - (1 - 1/q)^F]$.
Similarly, we denote as $\beta'_0$ the subset of
agents in $\beta$ that initially do not share any component with the state $y^f$. 

%%%%%%%%%%%%%%%%%%%%%%%%%%%%%%%%%%%%%%%%%%%%%%%%%%%%%%%%%%%%%%%%%%%%%%
\begin{figure}[h]
\begin{center}
\includegraphics[width=0.62\linewidth,angle=0]{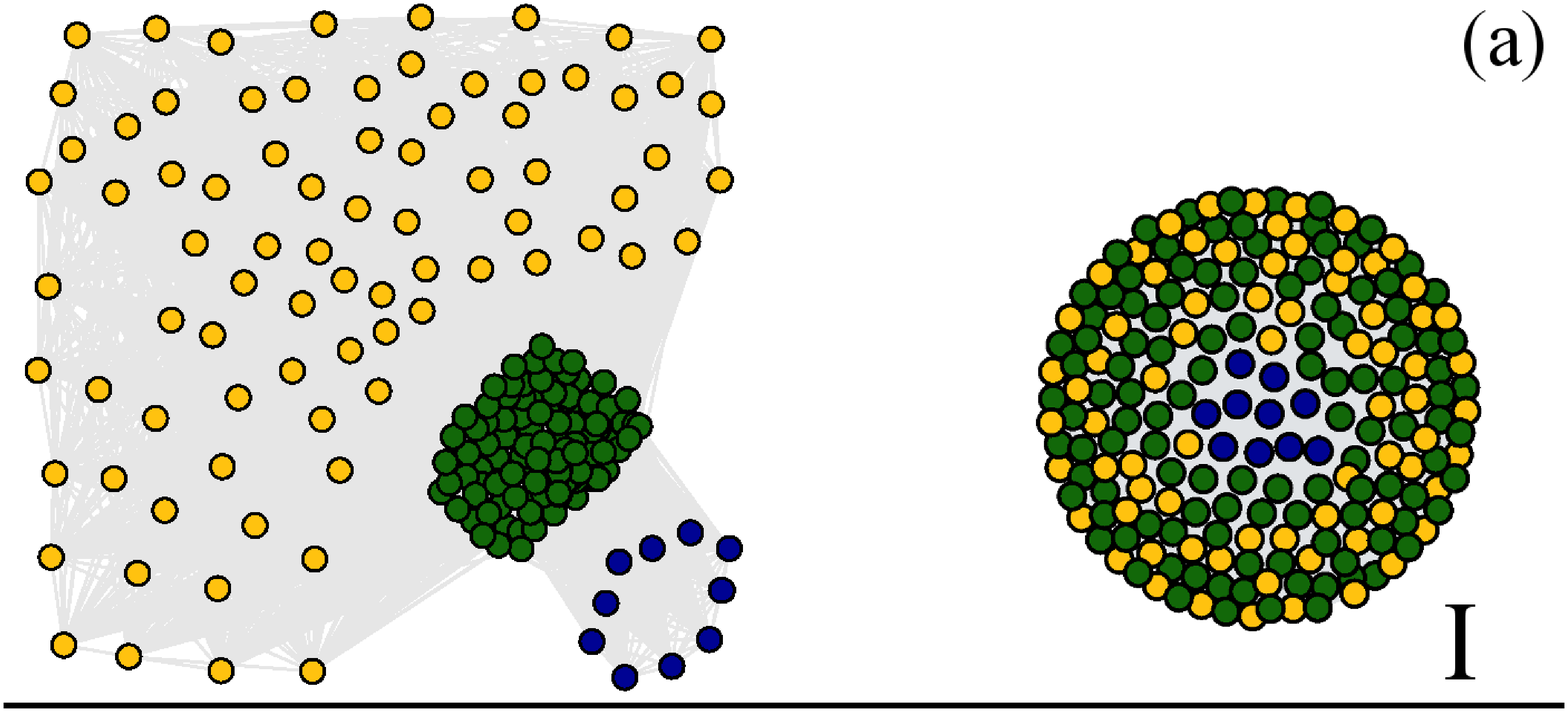} \\
\vspace{1mm}
\includegraphics[width=0.62\linewidth,angle=0]{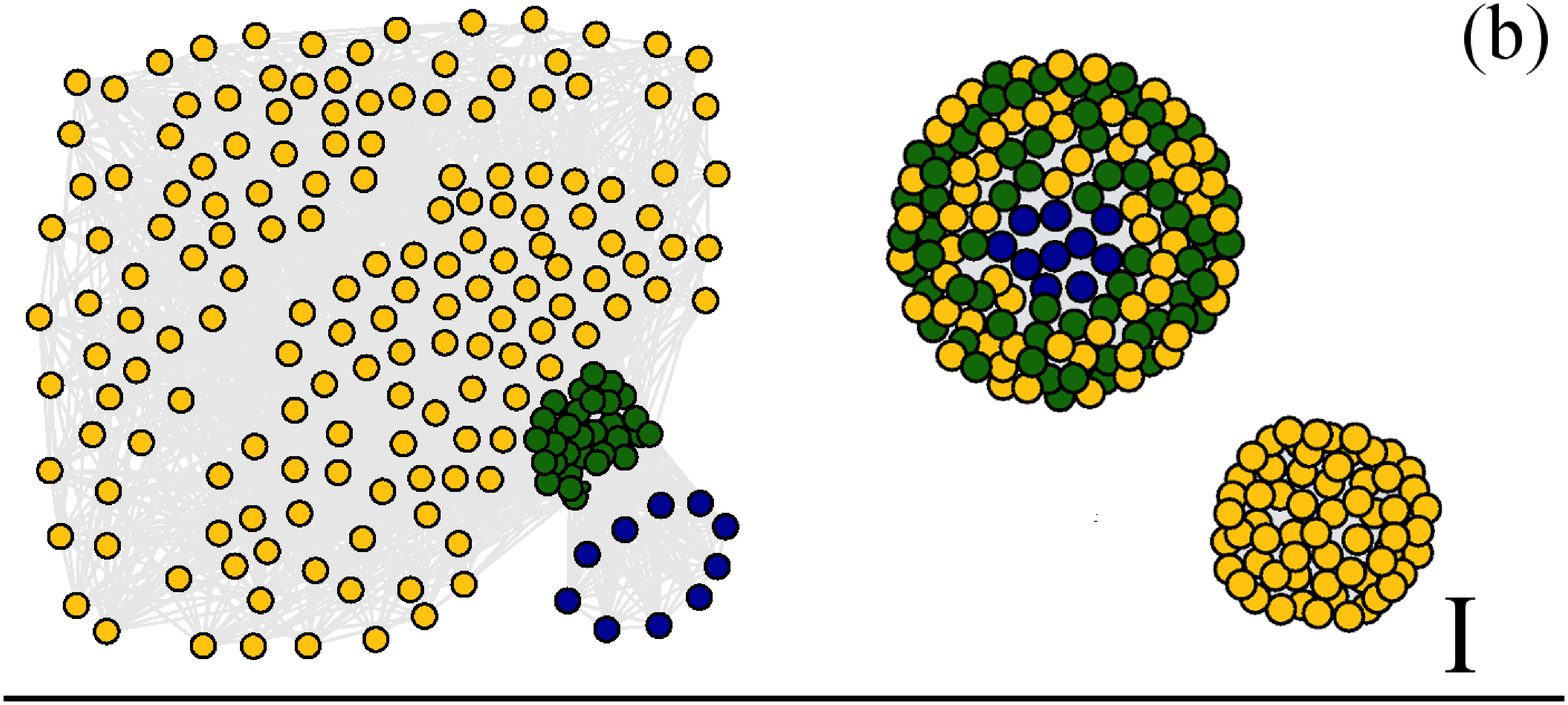} \\
\vspace{1mm}
\includegraphics[width=0.62\linewidth,angle=0]{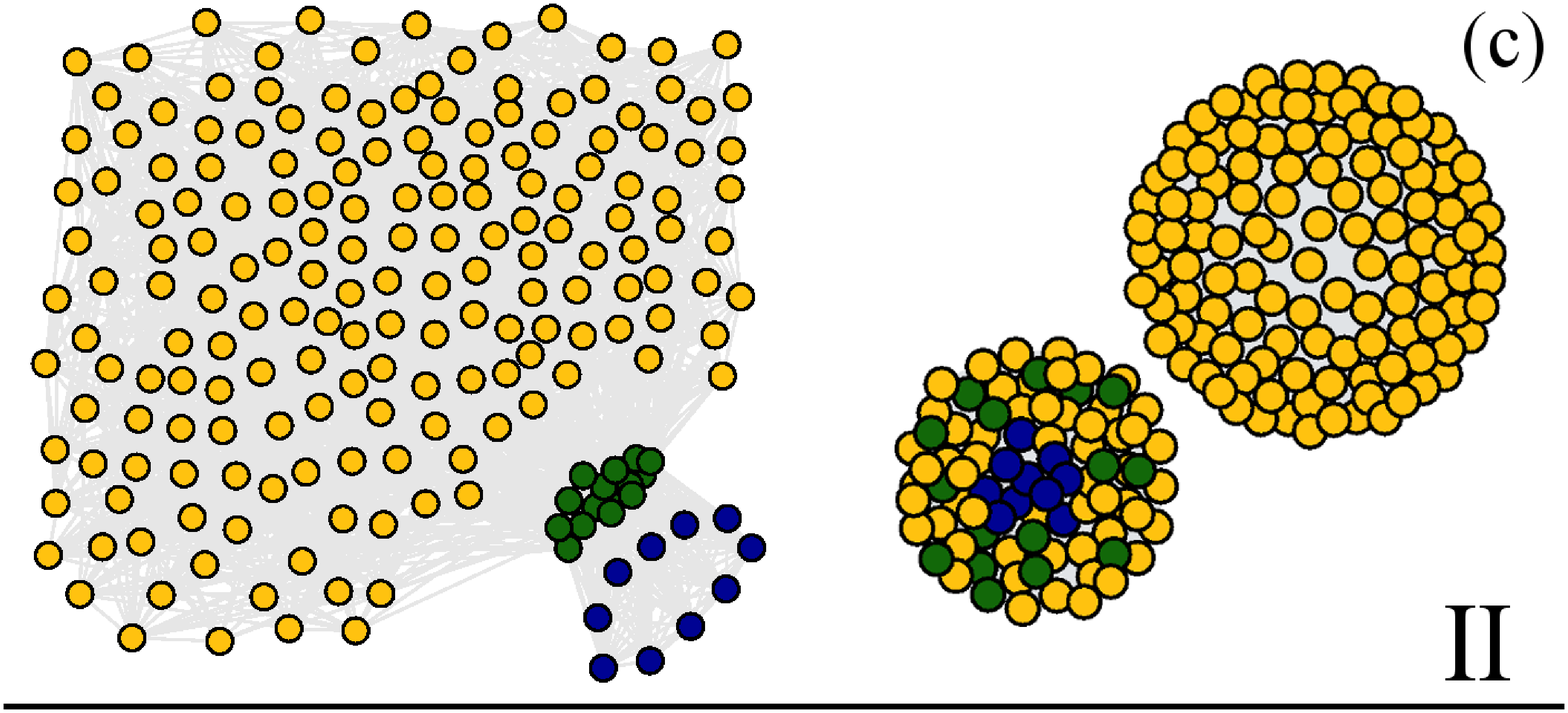} \\
\vspace{1mm}
\includegraphics[width=0.62\linewidth,angle=0]{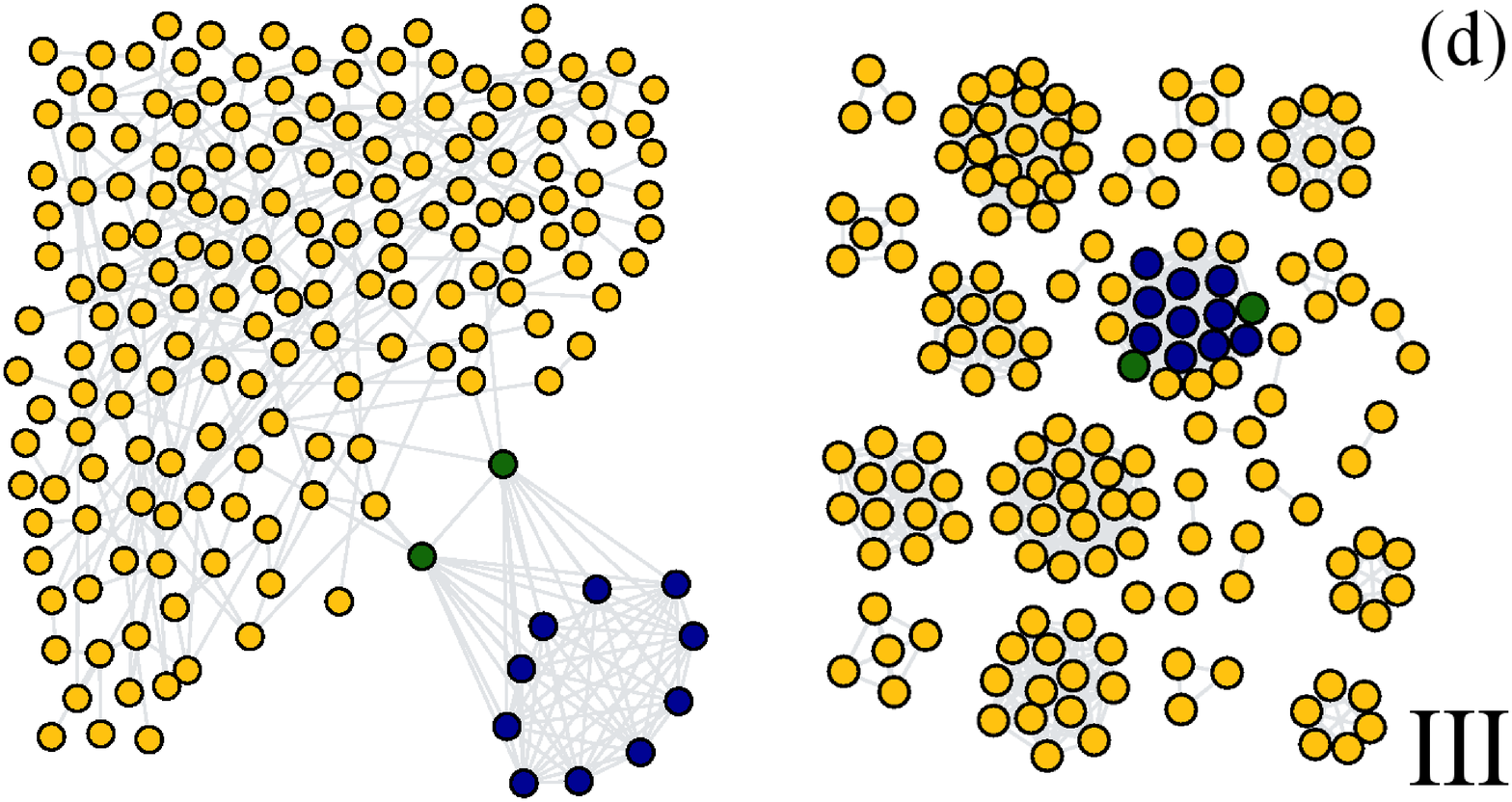}
\end{center}
\caption{Initial (left column) and final (right column) configurations of the network of interacting states, for different values of $q$, 
where a link (gray line) between any two agents exists if their states share at least one component. 
Opinion leaders (set $\alpha$) are depicted in blue color forming a circle; agents that initially 
share at least one component with the state of the opinion leaders (set $\beta_0$) are shown in green;
and agents that initially do not share any component with the state of opinion leaders (set $\beta'_0$) are drawn in yellow. 
In the final configurations (right column), a group of connected agents corresponds to a domain.
Fixed parameters are $N=200$, $F=5$, $\rho=0.05$. 
(a) $q=20$ (phase I, with $\sigma=0$, $S_\beta(y^f)=1$). (b) $q=30$ (phase I, with $\sigma=0$, $S_\beta(y^f)<1$). 
(c) $q=100$ (phase II, $\sigma>0$). (d) $q=800$ (phase III, $\sigma=0$, $S_\beta  \to 0$).}
\label{F4}      
\end{figure}
%%%%%%%%%%%%%%%%%%%%%%%%%%%%%%%%%%%%%%%%%%%%%%%%%%%%%%%%%%%%%%%%%%%%%%

Figure~(\ref{F4}) shows the initial (left column) and final (right column) configurations of the interaction network so defined, 
with a fixed fraction of opinion leaders $\rho$, and for different
values of the parameter $q$. Nodes representing opinion leaders in set $\alpha$, as well as agents in subsets $\beta_0$ and $\beta'_0$, 
are identified by respective colors. In the final configurations, a set of connected agents corresponds to a domain; 
although, for simplicity, the states associated to different domains are not distinguished by different colors.

Figures~(\ref{F4}a) and (\ref{F4}b) correspond to two realizations of phase I. For a small value of $q<q_*$ (Fig.~(\ref{F4}a)),
the size $N_{\beta_0}$ is large and both, the probability that the states of agents in  $\beta_0$ copy additional components of $y^f$ and
the probability that the states of agents in  $\beta'_0$ acquire components of $y^f$ through their interaction with agents in $\beta_0$,
are high. Eventually, all agents in $\beta$ end up sharing all components of $y^f$ and forming one large domain of size  $S_\beta(y^f)=1$.
As $q$ is increased (Fig.~(\ref{F4}b)), but still below the value $q_*$, those probabilities decrease and not 
all agents in subset $\beta'_0$ are able  
to acquire any component of $y^f$. As a consequence, the final interaction network becomes divided into two subgraphs containing agents from
$\beta_0$ and $\beta'_0$; the largest subgraph being the largest domain possessing the state $y^f$.

Figure~(\ref{F4}c) shows the configurations of the interaction network for  $q_* < q < q_c$. The final network 
also consists of two domains, but it is associated to phase II. Now the size of subset $\beta_0$ has become too small
to contribute efficiently to the transmission of the state of the opinion leaders  $y^f$ to agents in $\beta'_0$.
Correspondingly, the size of set $\beta'_0$ is large enough to allow a majority of its agents 
to form, through their interactions,
the largest domain in a state different from $y^f$. Finally, Fig.~(\ref{F4}d)
displays the initial and final aspects of the interaction network corresponding to phase III, for $q >q_c$. Since $q$ is large, 
there is a large number of states
available to agents in $\beta$ that are non-interacting with the state $y^f$. This situation leads to 
the formation of many small domains and the fragmentation of the network of interacting states.
Thus, Fig.~(\ref{F4}) illustrates how the evolution of the system towards its final phase depends on the initial configuration of the 
interaction network, determined by the size of the subset of agents $\beta_0$ whose states initially possess some overlap with
the state of the opinion leaders.

\vspace{-1mm}

\section{Local connectivity.}
To investigate the role of the local connectivity of the network on the appearance of phase II,
we next consider the dynamics of the system defined on a random network of $N$ nodes having average degree $\bar{k}$.
A fully connected network studied above corresponds to the case $\bar{k}=N-1$.

\begin{figure}[h]
\begin{center}
\includegraphics[width=0.43\linewidth,angle=90]{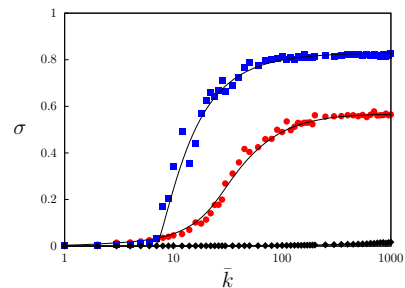}
\end{center}
\caption{Quantity $\sigma$ as a function of the average degree $\bar k$ of a random network  
for parameter values:
$q=50$ (circles),  
$q=100$ (squares), 
$q=2000$ (diamonds). 
Fixed parameters are $\rho=0.05$, $N=1000$, $F=5$.
Each data point is an average over $50$ independent realizations of initial conditions.}
\label{F5}      
\end{figure}

Figure~(\ref{F5}) shows the quantity $\sigma$ as a function of $\bar{k}$ for networks of size $N=1000$, for several values of $q$
with other parameters fixed. 
For values $q > q_c$, a disordered state with $\sigma \to 0$ (both $S_\beta \to 0$, $S_\beta(y^f) \to 0$) is reached for all values of
$\bar{k}$ in random networks. 
However, in connected random networks ($\bar{k} \geq 8$ for $N=1000$), for $q_* < q < q_c$ 
the largest domain in the network possesses a state orthogonal to $y^f$  for which $\sigma > 0$.
Then, global interactions are not essential for
the rise of a majority, alternative group in the presence the opinion leaders; rather this effect depends on the existence of
a minimum number of long range connections  
that is small compared to the size of the network. 

%%%%%%%%%%%%%%%%%%%%%%%%%%%%%%%%%%%%%%%%%%%%%%%%%%%%%%%%%%%%%%%%%%%%%%

\vspace{-4mm}

\section{Conclusions.}
We have investigated the effect of opinion leaders or influentials in the collective behavior of a social system, 
as well as the role of the network topology in this effect.
Our dynamical model is based in Axelrod's rules for cultural interaction among social agents that allow for non-interacting states.  
Opinion leaders have been characterized by their unidirectional influence on other agents, a behavior similar to that
of an external field or mass media acting on a system. In this sense, opinion leaders can be considered as distributed mass media, 
scattered advertisers, or as a nonuniform field. 

We have found three collective phases in the system depending on
parameter values: one ordered phase having the state imposed by the opinion leaders;  
another nontrivial ordered phase consisting of a large domain on a state orthogonal or alternative to that of the opinion leaders,
challenging the influentials hypothesis;
and a disordered phase.
We have shown that the resulting 
phase in the system is controlled by the size of 
the subset of agents whose states initially possess some overlap with the state of the opinion leaders. This explains why
a critical mass of agents susceptible to influence, rather than influentials, 
drives some propagation processes \cite{Watts}.

We have shown that 
the rise of an alternative, majority group in the presence the opinion leaders depends on the existence of
a minimum number of long-range connections in the 
network. Thus, this
phenomenon should be observable in experiments measuring influence, product adoption, or viral marketing in social networks, 
such as those performed with Facebook users \cite{Aral}.

Our results suggest that the emergence of a self-organized phase with a state different from
that of leaders should occur in other non-equilibrium systems possessing 
non-interacting states in their dynamics and enough long-range interactions in their underlying
network. This phenomenon could be expected in social and biological systems
able to exhibit clustering, aggregation and migration, whose dynamics often possess a threshold 
condition for interaction \cite{Deffuant,Mikhailov,Weis,Krause}. 
Future extensions of this model should include the consideration of diverse interaction rules, the competition 
of opinion leaders in different or variable states, and the role of complex network topologies. 

\section*{Acknowledgments}
%\begin{acknowledgments}
This work was supported by project No. C-1906-14-05-B from CDCHTA, 
Universidad de Los Andes, Venezuela.   
J.C.G-A acknowledges support from CNPq, Brazil, under project No. 150566/2015-8. 
M. G. C. is grateful  to the Senior Associates Program of
the Abdus Salam International Center for Theoretical Physics, Trieste, Italy. 
%\end{acknowledgments}

\end{document}